\documentclass{aa}

\usepackage{txfonts}
\usepackage{graphicx}
\usepackage{natbib}
\bibpunct{(}{)}{;}{a}{}{,}
 
\begin{document}
 
\title{L1521E: the first starless core with no molecular depletion}

\author{M. Tafalla 
\and
J. Santiago 
}
 
\institute{Observatorio Astron\'omico Nacional, Alfonso XII 3, E-28014 Madrid,
Spain
}
 
\offprints{M. Tafalla \email{m.tafalla@oan.es}}
\date{Received -- / Accepted -- }
 
\abstract{
L1521E seems unique among starless cores. 
It stands out in a distribution
of a ratio $(R)$ that we define to asses core 
evolution, and which compares the emission of the easily-depleted
C$^{18}$O molecule with that of the hard to deplete, late-time 
species N$_2$H$^+$.
While all cores we have studied so far have $R$ ratio lower than 1,
L1521E has an $R$ value of 3.4, which is 8 times the 
mean of the other cores. To understand this difference,
we have modeled 
the C$^{18}$O and N$_2$H$^+$ abundance profiles in L1521E
using a density distribution derived from 1.2mm continuum data.
Our model shows that the C$^{18}$O emission in this core is consistent with 
constant abundance, and this makes L1521E the first core with no C$^{18}$O
depletion. Our model also derives an unusually low N$_2$H$^+$ abundance.
These two chemical peculiarities suggest that L1521E has contracted to
its present density very recently, and it is therefore an extremely 
young starless core. Comparing our derived abundances with a chemical
model, we estimate a tentative age of $\le 1.5 \times 10^5$ yr, which is
too short for ambipolar diffusion models.
}

\authorrunning{Tafalla \& Santiago}
\titlerunning{L1521E: the first starless core with no molecular depletion}
 
\maketitle

\section{Introduction}

Observations of starless 
dense cores over the last few years have shown that molecular
depletion onto cold dust grains is a common feature 
in low-mass star-forming regions
(e.g., \citealt{cas99,taf02}). As a core contracts, the
amount of depletion of certain molecules like CS and CO 
increases with time (e.g., \citealt{ber97,aik03}), so we
can expect that depletion will provide an accurate clock for
core evolution when its details are well understood.
Even before this is a reality, depletion can be used as
a qualitative time marker, in
the sense that evolved cores should be more depleted of
certain species than younger cores. When we attempt to
build a sequence of cores having different amounts
of depletion, however, 
we find that the population of undepleted (or lightly)
depleted cores is missing. These undepleted cores should represent the 
earliest stages of core contraction, and probably 
they have not been identified so far because of an observational
bias. Given the interest of young cores for studies of core
contraction, finding a member of this
population is an urgent challenge.

To identify cores with low degree of depletion, we
have started a systematic study of cores 
with weak NH$_3$ lines in the survey of \citet{ben89},
as previous  observations of cores bright in NH$_3$ have only found cases 
of strong depletion (e.g., \citealt{cas99,taf02}).
Among our targets, the L1521E core in Taurus 
has revealed itself as a core with negligible 
depletion, and here we report 
a preliminary analysis of its properties.
This core has already been identified as a very young core 
by \citet{hir02}, who found it very prominent in 
carbon-chain molecules, comparable to TMC-1
(see also \citealt{aik03}). In this paper we 
show that L1521E has the lowest measured 
level of C$^{18}$O depletion, and in 
a forthcoming article we will present the results of an 
ongoing chemical survey of this core.

\section{Observations}

We observed L1521E in C$^{18}$O(1--0) and N$_2$H$^+$(1--0) 
with the FCRAO 14m telescope during 2002 March. The observations were
done with the SEQUOIA array in on-the-fly position switching mode.
The facility correlator provided a velocity resolution of about 
0.07 km s$^{-1}$, and all
data were converted into the main beam brightness scale assuming 
an efficiency of 0.55. The telescope beam size was approximately
$50''$.

We observed L1521E with the IRAM 30m telescope in 2003 September (lines)
and October (continuum). The line observations consisted of
maps in N$_2$H$^+$(1--0) and C$^{18}$O(2--1), and 
crosses in C$^{17}$O(1--0), C$^{17}$O(2--1), and C$^{34}$S(2--1). 
All data were taken
in frequency switching mode with  
a velocity resolution of 0.03 km s$^{-1}$, and were converted
into main beam brightness units
using standard efficiencies. The 1.2mm continuum observations
were done with MAMBO-2 in on the fly 
mode. Sky dips were used to measure
the atmospheric extinction, and observations of planets 
were used for calibration.
The resolution of the 30m telescope 
varies from $26''$ for N$_2$H$^+$(1--0)
to $11''$ for C$^{18}$O(2--1) and the 1.2mm continuum.

\section{The $R$ ratio and the uniqueness of L1521E}

\begin{figure*}
\centering
\resizebox{15cm}{!}{\includegraphics{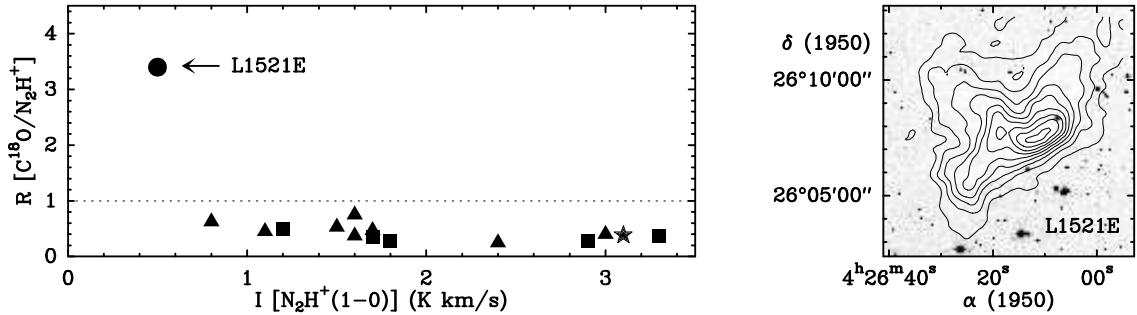}}
\caption{{\bf Left: \/} $R$ parameter 
(C$^{18}$O(1--0)/N$_2$H$^+$(1--0) ratio at the core center) 
as a function of the 
central N$_2$H$^+$(1--0) integrated intensity. Each point represents 
a core and the dashed line at $R=1$ represents the average 
$R$ value for the core models with no C$^{18}$O depletion in
\citet{taf02}. The squares represent cores studied in that paper 
(L1400K, L1517B, L1498, L1495, and L1544 by increasing N$_2$H$^+$ intensity), 
the triangles represent cores from an unpublished survey
(L981-1, L1197, L1155C1S, L1155C1N, L1512, L234E-1, L1333, and L694-2, 
same order), 
and the star symbol indicates B335, the only core with a YSO.
Note how L1521E stands out from the core sample. {\bf Right: \/}
Large scale C$^{18}$O(1--0) integrated intensity map overlayed on a
DSS blue image. Contours are .2, .4, ... K km s$^{-1}$. All radio data 
are from FCRAO.
\label{fig1}}
\end{figure*}

To search for cores at the earliest stages of contraction,
it is convenient to define a simple observable that measures
depletion without the need of a complicated radiative
transfer calculation. One possibility is to compare the emission of
two molecules known to have
different depletion behaviors at typical core densities, and for
this we choose
C$^{18}$O and N$_2$H$^+$. C$^{18}$O is a heavy depleter at densities 
of a few $10^4$ cm$^{-3}$, while N$_2$H$^+$ seems to remain in the gas phase
up to densities of about $10^6$ cm$^{-3}$ (e.g., \citealt{cas99,taf02}).
Thus, if we define $R\;[$C$^{18}$O/N$_2$H$^+]$ ($R$ hereafter) as 
the ratio between the integrated intensities
of C$^{18}$O and N$_2$H$^+$ at the center of a core, we expect
$R$ to be a sensitive indicator of depletion. In particular,
we expect $R$ to be low in cores with strong C$^{18}$O depletion and high
in cores with little C$^{18}$O depletion. 

Our definition of $R$ 
has the additional advantage that it compares 
two molecules with different gas-phase formation time scales:
C$^{18}$O forms relatively quickly in the core history while N$_2$H$^+$
is a late-time molecule (e.g., \citealt{aik03}). Thus, if chemical 
youth is correlated with
lack of depletion, the $R$ ratio should be further enhanced in young 
cores
because of their relatively lower N$_2$H$^+$ abundances. A signature
of youth and lack of depletion, therefore, is the combination of
a large $R$ and a relatively low N$_2$H$^+$ intensity.

To calibrate the $R$ ratio,  
we test its behavior in a sample
of cores selected for their strong N$_2$H$^+$ or NH$_3$ emission, and
therefore likely to be evolved. This sample includes the 5 
cores studied in \citet{taf02}, a set of 8 additional cores from an
unpublished survey, and a
core with an embedded YSO, B335. All these cores have been  mapped in 
C$^{18}$O and N$_2$H$^+$ with the FCRAO telescope, and 
their $R$ ratio has been estimated at the 
position of maximum N$_2$H$^+$ emission. In addition and as a 
reference, we have estimated the $R$ ratio for the 5 Monte Carlo core
models with constant C$^{18}$O abundance in \citet{taf02}, 
finding an average $R$ value of 1.0. 

As Figure 1 illustrates, all cores in our sample lie below the $R=1$
line despite their factor of 4 spread in N$_2$H$^+$(1--0) 
integrated intensity. The sample has a mean $R$ value of 0.43 with an rms 
of 0.14, and its location under the $R=1$ line is suggestive of strong 
C$^{18}$O depletion (a conclusion supported by detailed 
radiative transfer modeling of some objects). This
systematic behavior suggests that $R=1$ can be taken as a
reasonable boundary between cores with and without C$^{18}$O depletion, 
and that a search for undepleted cores should look for objects
above the $R=1$ line.  Thus, over the last few years we have carried 
out a search for high-$R$ cores using the FCRAO telescope,
mapping core candidates both in N$_2$H$^+$(1--0) and C$^{18}$O(1--0).
Although the search is still ongoing and several candidates seem promising,
one object stands out because it combines the highest $R$ value and the
lowest N$_2$H$^+$ intensity. This core, L1521E, shown in Fig. 1 as 
the outlying point labeled with its name, has $R=3.4$, which 
is 8 times larger than the mean or 20 times the rms. These characteristics
make L1521E
the best candidate we know of to an undepleted core, and as we will
in the following section, a detailed analysis of its molecular composition
confirms this interpretation.

\section{The internal structure of L1521E}

\begin{figure*}
\centering
\resizebox{16cm}{!}{\includegraphics{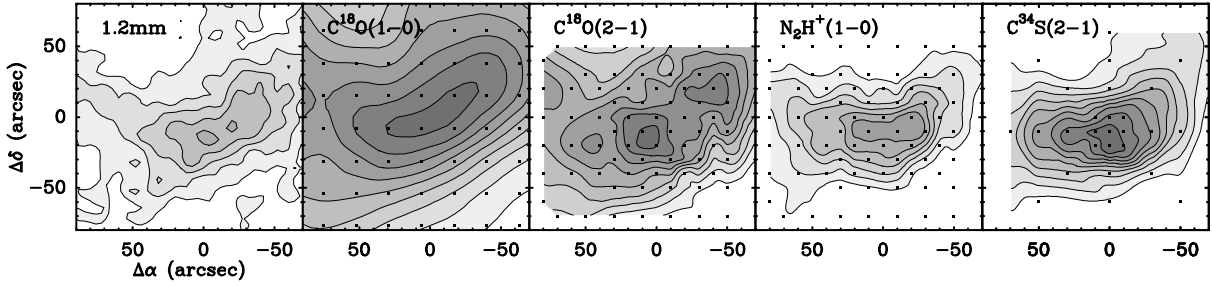}}
\caption{Maps of L1521E in 1.2mm continuum, C$^{18}$O(1--0), C$^{18}$O(2--1),
N$_2$H$^+$(1--0), and C$^{34}$S(2--1). Note the similar distribution of
all tracers, something not seen in cores with depletion. First 
contour and contour spacing are 4 mJy beam$^{-1}$ for the 1.2mm continuum map, 
0.2, 0.3, 0.1, and 0.1 K km s$^{-1}$ for C$^{18}$O(1--0), C$^{18}$O(2--1),
N$_2$H$^+$(1--0), and C$^{34}$S(2--1), respectively. Offsets are referred to
$\alpha_{2000}=4^{\rm h}29^{\rm m}15\fs7$, $\delta_{2000}=26\degr14'5''$.
Note that the C$^{34}$S(2--1) map is heavily undersampled. The C$^{18}$O(1--0)
map is from FCRAO data and the other maps are from IRAM 30m data.
\label{fig2}}
\end{figure*}

The high $R$ ratio of L1521E can arise from a high central C$^{18}$O abundance,
a low N$_2$H$^+$ fraction, a combination of both, or some excitation effect
(after all, the two molecules have very different dipole moments). To discern
between these options, we now solve the radiative transfer 
of the two species and determine their abundance profiles. For this,
we rely on observations of the 1.2mm continuum, C$^{18}$O(1--0),
C$^{18}$O(2--1), C$^{17}$O(1--0), C$^{17}$O(2--1), and N$_2$H$^+$(1--0), 
shown in part in Figure 2.

\subsection{Physical parameters}

We first derive the density profile of L1521E from
its distribution of the 1.2mm
continuum emission. This emission is optically thin and appears
unaffected by chemical differentiation or excitation effects, so
it can be easily inverted into a density  distribution
(e.g., \citealt{and96}). As Fig. 2 shows, the 1.2mm emission
(like that of the molecules) is elongated
NW-SE, and therefore deviates from spherical symmetry. Thus, to derive
a density distribution that can be used by our spherically-symmetric
Monte Carlo radiative transfer analysis (see
below), we first average elliptically the observed emission, assuming an
aspect ratio of 2:1 and a position angle of $-60^\circ$. Then, we find an
analytic
density profile that after simulating an on-the-fly observation
predicts the correct radial profile of 1.2mm continuum emission (see
\citealt{taf02} for details). For consistency with previous work,
we assume a dust emissivity $\kappa_{1.2mm}=0.005$ cm$^2$ g$^{-1}$
and a dust temperature of 10 K (see below). In this way,
we derive a density distribution of the form
$n(r)=n_0 /(1+(r/r_0)^\alpha),$ where $n_0=2.7 \times 10^5$ cm$^{-3}$, 
$r_0=6.3 \times 10^{16}$ cm ($=30''$ at 140 pc),
and $\alpha=2$. Fig. 3 shows that this profile provides a reasonable fit to
the observed 1.2mm continuum emission.

The NH$_3$ emission from L1521E is extremely weak \citep{ben89,hir02},
so we rely on C$^{17}$O 
to determine the gas kinetic temperature. C$^{17}$O seems undepleted at the 
core center (see below), and fits to its hyperfine structure indicate 
that the emission
is optically thin. Thus, we can convert the observed central
2--1/1--0 intensity ratio ($\approx 1.3$)
into an excitation temperature ($\approx 9$ K) which we approximate as 10 K.
The 2--1/1--0 line ratio seems constant with radius, so 
the excitation temperature must also be close to constant. Given the 
low dipole moment of C$^{17}$O (0.1 D), 
this excitation temperature should be a good approximation 
to the gas kinetic temperature, and we will assume in our modeling a 
constant gas temperature of 10 K. 

Finally, we characterize the velocity field of L1521E using hyperfine
fits to its N$_2$H$^+$(1--0) emission. These fits indicate an intrinsic
N$_2$H$^+$ linewidth of 0.3 km s$^{-1}$, 
which means a nonthermal component of about 0.27 km s$^{-1}$ (FWHM).
When modeling the C$^{18}$O(1--0) lines, a slightly larger
linewidth is needed in the outer core, and we have parametrized this
behavior as a linear increase in the non thermal component
up to 0.58 km s$^{-1}$ at $4 \times 10^{17}$ cm (the core outer radius
in our model).

\subsection{Molecular abundance profiles}

\begin{figure*}
\centering
\resizebox{16cm}{!}{\includegraphics{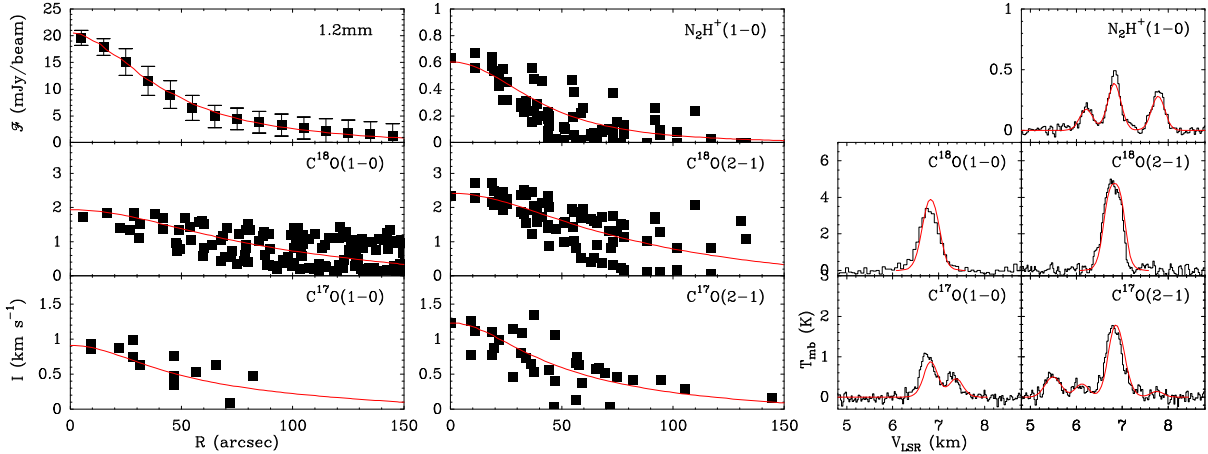}}
\caption{Radial profiles of integrated intensity (left) and central emerging
spectra (right) for 1.2mm continuum, N$_2$H$^+$(1--0), C$^{18}$O(1--0), 
C$^{18}$O(2--1), C$^{17}$O(1--0), and C$^{17}$O(2--1)
(data origin as in Fig. 2). The squares 
and histograms are observed data, and
the solid lines are model predictions. The model fitting the
1.2mm continuum is the basis of our core density profile, and the
models fitting the line emission have been calculated using a Monte Carlo
radiative transfer model. All models assume constant molecular abundance
(see text).
\label{fig3}}
\end{figure*}

We solve
the radiative transfer for N$_2$H$^+$ and the CO isotopomers
to derive their radial profiles of abundance. For this, we use
a Monte Carlo code \citep{ber79, taf02, taf04} and
fit simultaneously both the radial profiles
of integrated intensity and the emergent spectra of all transitions
(the radial profiles have been derived with the same elliptical
average as used for the continuum). Given the previous 
parameterization of the L1521E core, the only free parameters to fit the
data are the abundances of each species.

To determine the N$_2$H$^+$ abundance we use the 
1--0 transition. This transition has 7 hyperfine components,
and our model treats each component as a separate line, taking explicitly
into account the hyperfine splitting of the first 3 rotational levels
(see \citealt{taf04} for further details). We have
run this model assuming constant N$_2$H$^+$ abundance, and as Figure 3 shows,
the predicted emission for a value of $2 \times 10^{-11}$
fits simultaneously the radial
profile of integrated intensity and the central emerging spectrum.
A constant N$_2$H$^+$ abundance is also found in 
other cores, but the value we derive for L1521E is about 8 times
lower than that derived for the L1498 and L1517B cores using the
same code \citep{taf04}. This difference is highly significant
and makes L1521E the starless core with the lowest N$_2$H$^+$ abundance
(note that the N$_2$H$^+$ abundances estimated by \citealt{taf02} 
are artificially low because of their simplified treatment of the 
hyperfine structure).
The low N$_2$H$^+$ 
abundance explains the location of L1521E in the far left of Fig. 1.

To determine the abundance of the CO isotopomers, we
now fit the radial profiles and emergent
spectra of C$^{18}$O(1--0), C$^{18}$O(2--1), C$^{17}$O(1--0), and 
C$^{17}$O(2--1). We first
fix the C$^{18}$O/C$^{17}$O abundance ratio to its ISM value 
(3.65, \citealt{pen81}), so we are left with one abundance profile 
as free parameter.
In contrast with what is found in other cores, 
the L1521E data can be fit with a model of constant CO 
abundance. This is illustrated in Fig. 3 and makes L1521E 
the first core with no sign of CO depletion. Our best-fit abundance is
X(C$^{18}$O) = 0.7 $\times 10^{-7}$, 
very similar to that found in the outer parts of
L1498 and L1517B, although about 2.5 times lower than the
standard determination by \citet{fre82}. This lower value may 
indicate a slight error in the dust parameters
or that the \citet{fre82} abundances need revision. 
In any case, it is the combination of high CO and low N$_2$H$^+$
abundances what causes the hight $R$ ratio of L1521E.

Although an analysis of the CS emission in L1521E 
is out of the scope of this paper, we note that preliminary
models indicate that this species also has constant abundance
all the way to the core center (already suggested by the
C$^{34}$S(2--1) emission in Fig. 2). CO and CS, therefore, 
seem not depleted in L1521E.

\section{An extremely young core}

The absence of CO (and CS) depletion in L1521E 
seems not to result from
molecules finding it harder to freeze out in this core.
Its central density is typical, as well as is its
gas temperature of 10 K (the dust temperature cannot
be much higher given the observed central density). The
very low abundance of the late-time molecule N$_2$H$^+$ 
further suggests 
that the lack of depletion in L1521E results
from this core being less chemically processed than
others, as it has also been argued by \citet{hir02} and
\citet{aik03}. L1521E, in fact, combines the lowest C$^{18}$O depletion
and lowest N$_2$H$^+$ abundance, so it appears to be
the less processed and therefore youngest core known.

To estimate how young L1521E is, we 
compare our observations with the predictions from the
recent model of \citet{aik03}, who have studied the evolution of
a core undergoing Larson-Penston contraction taking into account both 
gas-phase and grain-surface chemical reactions. Although the exact form of
contraction is unclear, this model seems preferable because it covers the 
chemical evolution of the gas since time prior to molecular depletion
when it has an initial density of  $10^4$ cm$^{-3}$.
Thus, we have taken the model abundance profiles 
at the earliest time calculated by the authors ($1.52 \times 10^5$ yr, their
Fig. 2a) and predicted intensities for the conditions of
L1521E. In addition of having to multiply the predicted abundances
by $\approx 3$ to approximately fit the data, we find that for C$^{18}$O
the slight central drop predicted by the chemical model is barely consistent
with the data, and that the predicted 
central increase of N$_2$H$^+$ abundance is
already too pronounced (the central density at this
early time is close to our estimate). This indicates
that the model predicts more chemically processing at 
time $1.52 \times 10^5$ yr than observed, even for the relatively
fast Larson-Penston flow. With the caveat that the time estimate
depends sensitively on the assumed sticking probabilities, we conclude
that the comparison implies a contraction age of $\le 1.5 \times 10^5$ yr.

Our contraction age for L1521E seems 
consistent with the estimate by \citet{lee99}
that the typical lifetime of starless cores is 0.3-1.6 Myr,
as this core is younger than others.
This age, however,
seems incompatible with models of core formation via ambipolar 
diffusion, 
as they require evolutionary times of the order of 1 Myr in the most
favorable conditions (e.g., \citealt{cio01}). Thus, if L1521E has 
significantly contracted on a time scale as short as suggested by 
chemical models, 
a more dynamical picture seems needed. The collision of supersonic
flows, as proposed by the supersonic turbulence scenario (e.g., 
\citealt{mac03}), however, seems also inconsistent with the observations,
given that the turbulent component of the linewidth ($\sigma \approx 0.11$ 
km s$^{-1}$) is still subsonic, and that the velocity differences
across the L1521E core are only of the order 
of the sound speed (a description of the core velocity structure 
will presented in a future paper). Clearly further analysis of L1521E
is needed to understand its apparent rapid formation and lack of 
molecular depletion. At this point, it is suggestive to speculate 
that L1521E belongs to an up to now overlooked population of extremely 
young cores.

\acknowledgements We thank the staffs of FCRAO and IRAM for help during the
observations and Yuri Aikawa for advice on chemical models. 
MT acknowledges support from grant AYA2000-0927 of the Spanish DGES.
The DSS
was produced at the Space Telescope Science Institute under US Government
grant NAG W-2166.

\bibliographystyle{apj}

\end{document}